\documentstyle[preprint,aps]{revtex}
\begin{document}

\title{
Magnetoelastic mechanism of spin-reorientation transitions
at step-edges.\\
}
\author{
A. B. Shick$^a$ \footnote{
Current address: Department of Physics,
University of California, Davis, CA 95616.}
, Yu. N. Gornostyrev$^{a,b}$ and A. J. Freeman$^a$,\\
$^a$Department of Physics and Astronomy,\\
Northwestern University, Evanston, IL 60208.\\
$^b$ Institute of Metal Physics, 620219 Ekaterinburg, Russia.}
\date{\today}
\maketitle

\begin{abstract}

The symmetry-induced magnetic anisotropy due to monoatomic steps at
strained Ni films
is determined using results of first - principles relativistic full-potential
linearized  augmented plane wave (FLAPW) calculations and an analogy
with the N\'eel model.  We show that there is a magnetoelastic
anisotropy contribution to the uniaxial magnetic
anisotropy energy in the vicinal plane of a stepped surface.
In addition to the known spin-direction reorientation transition
at a flat Ni/Cu(001) surface, we propose a spin-direction reorientation
transition
in the vicinal plane for a stepped Ni/Cu surface due to the magnetoelastic
anisotropy. 
We show that with an increase of Ni film thickness, the magnetization in
the vicinal plane turns perpendicular to the step edge 
at a critical thickness calculated to be in the range of 16-24 Ni layers for
the Ni/Cu(1,1,13) stepped surface. \\

\end{abstract}

Ni/Cu(001) is known as a unique system to show an in-plane to out-of-plane
spin-reorientation transition with an increase of Ni-overlayer thickness
\cite{Baberschke2,Bochi}. This is likely due to the positive magnetoelastic
energy contribution to the magnetic anisotropy energy (MAE),
caused by tetragonal strain in Ni-films due to
the film/substrate lattice mismatch \cite{Baberschke2,Wu,Hjor}.

We have investigated the uniaxial MAE for flat and stepped
Ni films, and use the results of first-principles MAE calculations 
for the strained Ni-films to estimate
the in-plane to out-of-plane spin reorientation transition thickness. 
Further, we also use an analogy with the N\'eel model with parameters
chosen from our first-principles calculations to estimate the
step-induced contribution to the MAE and to discuss the possibility of
a spin-reorientation transition at the Ni/Cu(001) monoatomic step.

The relativistic self-consistent version \cite{Shick} of the
full-potential linearized augmented plane-wave (FLAPW) method \cite{flapw}
is employed to perform self-consistent calculations for 
strained Ni films and bulk Ni with an (001) magnetization direction.
The local force theorem for magnetization rotation is then used for 
the MAE calculations \cite{MAE}.

For strained bulk Ni, we choose the bct (001) plane lattice constant
as 4.781 a.u. (equal to the Cu substrate)
and the (001) interlayer distance as 3.222 a.u.
in accordance with the
experimental strained structure of a Ni magnetic overlayer
on Cu(001) \cite{Baberschke1}.
Here,  343 k-points in 1/8th of the tetragonal 3D BZ  
were used for the self-consistent calculations.
Then 1728 k-points in 1/8th of the 3D BZ (13824 in the full BZ)
are used for the MAE calculations.
The MAE is shown in Table \ref{tab1}.
There is a very good agreement between our results and results of previous
perturbative state-tracking FLAPW calculations \cite{Wu} and relativistic full-potential
linear muffin-tin orbital (FLMTO) calculations
\cite{Hjor}. 

For Ni-films with a thickness (d) of 3,5,7,9,11 layers we choose a slab geometry with assumed
experimental strained structure of Ni magnetic overlayer on Cu(001) which is the same as in the
strained bulk Ni calculations.
Here, 100 k-points in 1/4th of the 2D BZ (400 k-points in the full 2D BZ) 
were used for the self-consistent calculations and
1600 k-points (6400 k-points in the FBZ) in
1/4 of the 2D BZ were used for the MAE calculations.


The calculated MAE as a function of Ni-film thickness is shown in Fig.1.
There is a very good linear dependence of the MAE on film thickness.
The uniaxial MAE of d magnetic bct(001) layers 
in the [(100),(010),(001)] coordinate frame (denoted by V) is then approximated by

\begin{equation}
\label{eq:MAE}
E \; = - \; d*K_v*V_z^2 - 2*K_s*V_z^2
\end{equation}

\noindent
where, $K_v$ is a ``volume" MAE per atom, $K_s$ is a ``surface" MAE per atom,
and $V_z$ is the magnetization direction cosine with respect to the z(V)-axis.
By fitting the calculated MAE (cf., Fig. 1) by Eq.(\ref{eq:MAE}), we obtain 
the ``volume" $K_v \; = \; 83.5 \; \mu eV$ and ``surface" $K_s \; = \; -446.5 \; \mu eV$ 
contributions. The calculated ``volume" contribution is in very good quantitative
agreement with the calculated uniaxial MAE for strained bulk-Ni (cf., Table {\ref{tab1}}).
This is  clear numerical evidence that 
the ``volume" contribution to the MAE in strained Ni films is the same as the MAE
of strained bulk Ni, which is qualitatively consistent with
N\'eel model arguments \cite{Chikazumi,Ohandley} and supports quantitatively an analysis
performed in Refs. \cite{Baberschke2,Hjor}.

Hjortstram {\it et al.} \cite{Hjor} have calculated the uniaxial MAE and magnetostriction
in strained bulk Ni and found good agreement between the
linear magnetoelastic theory and the results of first-principles
calculations for strain-induced MAE in the limit of small strains.
They also found that for assumed experimental strained structure
of Ni films  the MAE value for strained bulk Ni agrees well with
measured ``volume'' contribution to the MAE for Ni films on Cu(001) \cite{Baberschke2}.
The present calculations for both strained bulk Ni and Ni films (cf., Table I)
agree well with the results of \cite{Hjor}. The fact, that ``volume''
MAE (85 $\mu eV$) is somewhat bigger than the MAE  for strained bulk Ni
(57 $\mu eV$) can be due to computational differences and neglect of the
higher order terms in MAE expression Eq.(1) (which are small but not necessarily negligible).
Taking into account relatively small calculated MAE values we argue that both
the ``volume'' MAE for Ni films and MAE for strained Ni bulk agree surprisingly well.
This good numerical agreement makes it possible to conclude that ``volume'' MAE in Ni films
is due to the strain induced by film/substrate lattice mismatch.
This confirms quantitatively
conclusions of Hjortstram {\it et al.} \cite{Hjor} based on comparison between experimental
data for flat Ni films and the results of strained bulk Ni calculations.

Finally, using Eq.(\ref{eq:MAE}) and the calculated MAE data (cf., Table \ref{tab2}),
we calculate the in-plane to out-of-plane  spin-reorientation transition
thickness as $d_c = 11$ ML, which agrees qualitatively with the experimentally
observed spin-reorientation transition thickness of 7 ML for Ni/Cu(001).
(The demagnetization energy is omitted here because it is very small for Ni \cite{dd}.)
It is necessary to mention that the use of free-standing Ni-films to analyze
the MAE for Ni/Cu(001) is rather qualitative since we do not take into account
the interface contribution to the MAE and
the possible oscillations of the MAE with Ni-film thickness at the Ni/Cu(001) interface \cite{Wu}.
Moreover, the chosen geometry is an ``average" experimental
geometry and deviations due to relaxation are likely
to occur in the real interface.


To proceed further, let us now assume the N\'eel pair - interaction form
of the magnetic anisotropy energy ($E_{AB}$) \cite{Chikazumi}
for the pair of atoms A and B connected by
${\bf R}_{AB} = |{\bf R}_{AB}| (\beta_x,\beta_y,\beta_z)$,
with magnetic moment direction ${\bf m} = (V_x,V_y,V_z)$, where $\beta,V$
are the direction cosines; $L_{AB}$ is a N\'eel coefficient 
(the isotropic term is omitted). Hence,

\begin{eqnarray}
\label{eq:ma}
E_{AB} &=& -\;L_{AB} (\sum_{i=x,y,z} \beta_i V_i)^2 
\end{eqnarray}

The ``volume" and ``surface" MAE contributions in Eq.(\ref{eq:MAE}) are then expressed
as:

\begin{eqnarray}
\label{eq:volume}
K_{v} & = & + L (6 \beta_z^2 - 3)
\end{eqnarray}

\begin{eqnarray}
\label{eq:surface}
K_{s} & = &- L (3 \beta_z^2 - 1)
\end{eqnarray}

\noindent where, following Victora {\it et al.} \cite{Victora},
we consider only nearest-neighbor interactions and
do not take into account the change of the N\'eel parameter
due to the strain (which seems to be reasonable assumption in the limit of small strains, considered
here).

The value of the N\'eel parameter, $L = -396 \; \mu eV$, is then calculated
using Eq.(\ref{eq:volume}), the calculated value of the ``volume" anisotropy for strained bulk-Ni
(cf., Table {\ref{tab1}}) and $\beta_z^2 = 0.476$. The value of the ``surface" anisotropy is
then calculated using Eq.(\ref{eq:surface}) and is 169.5 $\mu eV$.
A comparison with the result
of first-principle calculations (cf., Table \ref{tab1})
clearly shows that this ``isotropic" variant of the N\'eel model 
does not even reproduce the sign calculated from the first-principles ``surface" MAE.

It is known \cite{Hjor2} that there is a significant change in the electronic
structure at the Ni-surface compared to the bulk. This change is especially pronounced
for the ``surface" layer and leads to significant modifications of the electronic
density of states near the Fermi level for the spin-minority band. This should lead to
a significant difference in ``in-plane" and ``inter-plane" N\'eel parameters for the
surface layer. Therefore, it is quite natural to introduce an additional in-plane N\'eel parameter
$M$ to describe the pair-interaction, Eq.(\ref{eq:ma}), within the surface layer.
This gives the following parametrization for the ``surface" anisotropy:

\begin{eqnarray}
\label{eq:surface2}
K_{s} & = &- L (3 \beta_z^2 - 1) \; + \; (L-M)
\end{eqnarray}

Then, we  approximate the MAE of a strained Ni-film by
choosing the N\'eel parameter $L$ to fit the ``volume" contribution to the MAE, Eq.(\ref{eq:volume}),
and then determine the in-plane parameter $M$ to fit the ``surface" contribution, Eq.(\ref{eq:surface2}).
Using the calculated values of
the ``volume" and ``surface" anisotropies
for the strained Ni-film (cf. Table \ref{tab1}), we obtain the
following values for the
N\'eel parameters, $L \; = \; -580 \; \mu eV$ and $M \; = \; 115 \; \mu eV$.

As  shown in Refs.\cite{Ohandley,Qiu}, the anisotropy due to the step along (110)
on the fct(001) surface can be written in terms of the contributions from ``step - corner"
and ``step - edge" atoms (we use the bct-system and  a fct (110) step is a bct (100) step): 

\begin{eqnarray}
\label{eq:corner}
E_{st} &= - K_{se} V_x^2 - K^1_{sc} V_z^2 - K^2_{sc} V_x V_z 
\end{eqnarray}

\noindent
where, $K_{se}$ is a ``step-edge" atom's MAE, and $K^1_{sc}$ and $K^2_{sc}$ are
a ``step-corner" atom's MAE.

Within the assumption of only nearest-neighbor interactions one can express
$K_{se}, K^1_{sc}$ and $K^2_{sc}$ in terms of the N\'eel parameters $L$ and $M$:


\begin{eqnarray}
\label{eq:corner2}
K^1_{sc} & = & 0.5 L (3 \beta_z^2 - 1) \; \nonumber \\
K^2_{sc} & = & - L \beta_x \beta_z  \;  \nonumber\\
K_{se}   & = & -0.5 M 
\end{eqnarray}

\noindent The values of $K_{se}, K^1_{sc}$ and $K^2_{sc}$ are then calculated for the Ni-
step using Eq.(\ref{eq:corner2}) and are shown in Table \ref{tab2}
together with values of the ``volume" and ``surface" anisotropies.

The symmetry induced uniaxial one-step MAE (per one atom in the xy-plane)
is then given \cite{difference} by:

\begin{eqnarray}
\label{eq:MAS}
E \; = \; - D*d K_v V_z^2 - 2*D*K_s V_z^2 - K_{se} V_x^2 \\ \nonumber
- K^1_{sc} V_z^2 - K^2_{sc} V_x V_z
\end{eqnarray}
where D is the number of one-step atoms (along x). We consider only the 
``surface" step and neglect the ``interface" step. As follows from Eq.({\ref{eq:MAS}),
the direction of the magnetization in the xy-plane is determined by the sign of $K_{se}$. In the
case of a Ni stepped surface, the magnetization is directed along the step-edge
(perpendicular to the x-axis) due to the negative sign of $K_{se}$ (cf. Table \ref{tab2}).
This is in qualitative agreement with experimental data \cite{Dhesi} for thin Ni-films
on stepped Cu(001).

Next,
we transform Eq.(\ref{eq:MAS}) to a vicinal coordinate frame (U)
using the relation between the old (V) and new (U) magnetization direction cosines
in terms of the vicinal angle $\alpha$. Keeping terms up to $\alpha^2$, and with

\begin{eqnarray}
\label{eq:frame}
V_x = (1 - 0.5\alpha^2)U_x + \alpha U_z\\ \nonumber
V_z = (1 - 0.5\alpha^2)U_z - \alpha U_x
\end{eqnarray}

we come to the following expression for the MAE in the xy-vicinal plane 

\begin{eqnarray}
\label{eq:MASxy}
E^{U}_{xy} \; = \;( -D*d K_v \alpha^2 - 2D*K_s \alpha^2 - K_{se} (1 - \alpha^2) \\ \nonumber
- K^1_{sc} \alpha^2 + K^2_{sc} \alpha)\; U_x^2 
\end{eqnarray}

\noindent Using the relation between $\alpha$ and $D$ ($\alpha \approx l_{op}/(a_{ip}D) \equiv x/D$)
and neglecting $1/D^2$-order terms we transform Eq.(\ref{eq:MASxy}) into:

\begin{eqnarray}
\label{eq:MASxy2}
E^{U}_{xy} \;=\;(- d \; x^2/D \; K_v  - 2 x^2/D \; K_s  - K_{se} \\ \nonumber
+ x/D K^2_{sc}) \; U_x^2
\end{eqnarray}

We next apply Eq. (\ref{eq:MASxy2}), with the parameters determined within the framework
of the modified N\'eel model (cf., Table \ref{tab2}), to calculate the MAE at the (1,1,13)
stepped surface.
Using the strain
parameter $x=0.674$ and step lenght $D=6$, we obtain the following contributions
to the MAE:

(1) The contribution due to the ``volume" magnetoelastic anisotropy is $-6.322 \times d \; \mu eV$;
it turns the magnetization {\it perpendicular} to the step and increases with an increase
of the film thickness;

(2) The contribution due to the ``surface" MAE is $68 \; \mu eV$;
it turns the magnetization {\it parallel} to the step and does not depend on the film thickness;

(3) The contribution due to the ``step-edge" MAE is $57.5 \; \mu eV$;
it turns magnetization {\it parallel} to the step and does not depend on the film thickness;

(4) The contribution due to ``step-corner" MAE is $+22 \; \mu eV$, and 
it turns the magnetization {\it parallel} to the step and does not depend on the film thickness;

Finally, for thicknesses below 24 ML, the contributions (2),(3) and (4) will
keep the magnetization parallel to the step in the vicinal (xy)-plane and the critical thickness
for the magnetization to turn perpendicular to the step edge is then estimated to be 24 ML.


In conclusion, Eq.(\ref{eq:MASxy2}) shows that there is a ``volume"
uniaxial magnetoelastic anisotropy
contribution to the symmetry-induced magnetic anisotropy in the vicinal plane of the stepped surface,
and it is caused by tetragonal strain in the magnetic film due to film/substrate lattice mismatch.
For monoatomic steps on Ni-films, this magnetoelastic contribution leads to an 
additional spin-reorientation transition in which the magnetization turns from parallel to 
perpendicular to the step alignment with increase of the Ni-film thickness. Using the results
of first-principles calculations for the strained Ni film 
and a N\'eel model-type analysis, we estimate the critical Ni-film thickness for this
transition to be 24 ML for the (1,1,13) stepped surface.
One has to take into account the difference between 
the surface  anisotropy in Ni/Cu(001) films and in the calculated free-standing Ni-films:
if we use the experimentally derived (at zero temperature) ``surface" MAE \cite{Baberschke2}
(i.e. -250 $\mu eV$),
we expect a critical thickness of 16 ML for this transition. This result shows the uncertainty 
and range of values possible at this level of theory.

We thank S. S. Dhesi and H. A. D\"urr for providing experimental results prior to 
publication and 
S. D. Bader, B. Ujfallysi, P. Weinberger, V. Drchal, and R.Q. Wu
for helpful discussions.

Work supported by the U.S. Office of Naval Research ( Grant No. N00014-94-1-0030)
and by a grant of computing time at the Navo Supercomputing Center.

\newpage

\begin{table}
\caption
{Uniaxial MAE of tetragonal strained bulk Ni ($K_v$), 
``volume" ($K_v$) and ``surface" ($K_s$) contributions to the MAE 
for strained Ni-films. All data are given in $\mu eV$.
}
\begin{tabular}{lllllll}
       &&This work & Exp. $^a$ & FLAPW $^b$ & FLMTO $^c$ \\
\hline
Bulk-Ni&$K_v$&57  &           & 65         & 60          \\
\hline
Ni-film&$K_v$&83.5& 30 (60-70)&      \\
       &$K_s$&-446.5&                                               \\
\end{tabular}
$^a$ Ref. \cite{Baberschke2} ( the data obtained for 300 K, 
in parentheses, are the result of a linear extrapolation of the temperature
dependent MAE to T=0 K)\\
$^b$ Ref. \cite{Wu}, $^c$ Ref. \cite{Hjor}.
\label{tab1}
\end{table}

\begin{table}
\caption
{
Values of $K_v$, $K_s$, $K_{se}, K^1_{sc}$ and $K^2_{sc}$ (in $\mu eV$) for the Ni-step.
}
\begin{tabular}{lllllll}
             &$K_v$& $K_s$ & $K_{se}$ & $K^1_{sc}$ & $K^2_{sc}$ \\
\hline
             &+83.5& -446.5& -57.5    & -124       & 195        \\  
\end{tabular}
\label{tab2}
\end{table}

\newpage

\centerline{FIGURES\\}

\vspace{0.5cm}

Fig. 1 The MAE (in $\mu \; eV$) for the strained Ni-films as a function of
the film thickness d (in ML).}
 

\begin{references}

\bibitem{Baberschke2} B. Schulz and K. Baberschke, Phys. Rev. {\bf B 50},
13 467 (1994); M. Farle {\it et al.}, Phys. Rev.{\bf B 55}, 3708 (1996).

\bibitem{Bochi} G. Bochi {\it et al.}, Phys. Rev. {\bf B 52},
7311 (1994).

\bibitem{Wu} R. Q. Wu, L. Chen, and A. J. Freeman, J. Appl. Phys. {\bf 81} 4417 (1997).

\bibitem{Hjor} O. Hjortstam, K. Baberschke, J.M. Wills, B. Johansson and O. Eriksson,
Phys. Rev. {\bf B 55}, 15026 (1997).

\bibitem{Shick} A.B. Shick, D.L. Novikov, and A. J. Freeman,
Phys. Rev. {\bf B 56}, R14259 (1997).

\bibitem{flapw} E. Wimmer, H. Krakauer, M. Weinert, and A. J. Freeman,
Phys. Rev. {\bf B 24}, 864 (1981).

\bibitem{MAE} MAE is calculated as a difference in one-electron energies for the 
(100) and (001) magnetization directions.


\bibitem{Baberschke1} K. Baberschke, J. Appl. Phys. {\bf A 62}, 417 (1996).

\bibitem{dd} 
Using the bulk-Ni magnetic moment value (0.61 $\mu_B$) and the
expression of Ref. \cite{Weinberger2}, we obtain
the demagnetization energy of -10 $\mu eV/atom$,
in very close agreement with the experimental value, -7.5 $\mu eV$
\cite{Baberschke2}.

\bibitem{Weinberger2} L. Szunyogh, B. Ujfalussy, P. Weinberger,
Phys. Rev. {\bf B 51}, 9552 (1995).

\bibitem{Victora}
R. H. Victora and J. M. MacLaren, Phys. Rev. {\bf B 47}, 11538
(1993).

\bibitem{Dhesi} S. S. Dhesi, H. A. D\"urr and G. van der Laan,
Phys. Rev. {\bf B 59}, 8408 (1999). 

\bibitem{Hjor2} O. Hjortstam {\it et al.}, Phys. Rev. {\bf B 53}, 9204 (1996),
and references therein.

\bibitem{Qiu} R. K. Kawakami, E.J. Escorcia-Aparicio, and Z.Q. Qiu,
Phys.Rev. Lett. {\bf 77}, 2570 (1996).


\bibitem{Chikazumi}
S. Chikazumi, {\it Physics of Magnetism}, (Krieger, Malabar, FL, 1986).

\bibitem{Ohandley}
D.S. Chuang, C.A. Ballentine, and R. C. O'Handley,
Phys. Rev. {\bf B 49}, 15084 (1994).

\bibitem{difference} The expression Eq. (\ref{eq:MASxy}) differs from 
Eq. (3) of Ref. \cite{Qiu} due to an additional uniaxial volume magnetoelastic 
term.



\end{references}
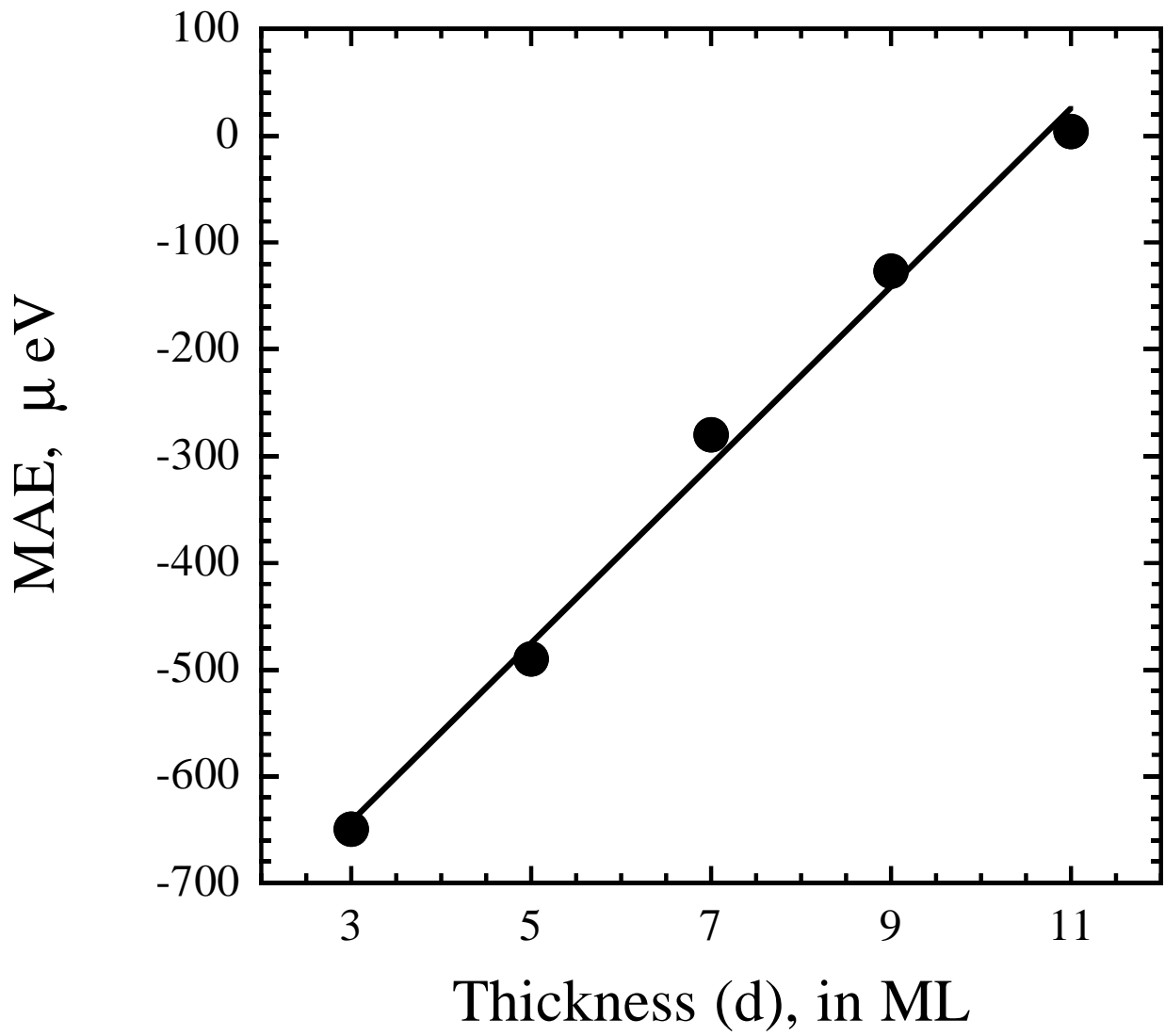
\end{document}